\documentclass[twocolumn,aps,prl,nobibnotes,groupedaddress]{revtex4-1}

\usepackage{graphicx}  
\usepackage{bm}        
\usepackage{amssymb}   
\usepackage{color}
\usepackage{amsmath}
\usepackage{braket}

\begin{document}

\title{A quantum transducer using a parametric driven-dissipative phase transition}

\author{Toni L. Heugel}
\affiliation{Institute for Theoretical Physics, ETH Zurich, 8093 Z\"urich, Switzerland}
\author{Matteo Biondi}
\affiliation{Institute for Theoretical Physics, ETH Zurich, 8093 Z\"urich, Switzerland}
\author{Oded Zilberberg}
\affiliation{Institute for Theoretical Physics, ETH Zurich, 8093 Z\"urich, Switzerland}
\author{R. Chitra}
\affiliation{Institute for Theoretical Physics, ETH Zurich, 8093 Z\"urich, Switzerland}

\begin{abstract}

We study a dissipative Kerr-resonator subject to both single- and two-photon detuned drives. Beyond a critical detuning threshold, the Kerr resonator exhibits a semiclassical first-order dissipative phase transition between two different steady-states, that are characterized by a $\pi$ phase switch of the cavity field. This transition is shown to persist deep into the quantum limit of low photon numbers. Remarkably, the detuning frequency at which this transition occurs depends almost-linearly on the amplitude of the single-photon drive. Based on this phase switching feature, we devise a sensitive quantum transducer that translates the observed frequency of the parametric quantum phase transition to the detected single-photon amplitude signal. The effects of noise and temperature on the corresponding sensing protocol are addressed and a realistic circuit-QED implementation is discussed.
\end{abstract}
\maketitle

\paragraph*{Introduction.}
Phase transitions  are commonly associated with  strongly-enhanced susceptibilities.  Proximity to phase transitions, therefore,  renders   systems highly sensitive to external perturbations.  Harnessing   this  augmented sensitivity for sensing and metrology using quantum systems~\cite{Degen2017} has  been the focus of numerous recent efforts in diverse settings, e.g., equilibrium systems~\cite{Zanardi2008}, PT-symmetric cavities~\cite{Liu2016}, dynamical phase transitions~\cite{Macieszczak2016}, and lasers~\cite{Lorenzo2017}. 
From this perspective, quantum driven-dissipative systems offer a fertile platform to devise such rich sensing protocols.  These systems are at the avantgarde of contemporary research at the interface between condensed matter physics and quantum optics~\cite{Hartmann2016,Noh2016}.
The dynamics of these intrinsically nonequilibrium systems is richer than that of  their equilibrium counterparts,  and dissipative phase transitions between different out-of-equilibirum  phases  can be controllably tuned.  Dissipative phase transitions  can be realized  in  various platforms, including cold atoms~\cite{Ritsch2013}, trapped ions~\cite{Blatt2012}, superconducting circuits~\cite{Schmidt2013}, and exciton-polariton cavities~\cite{Carusotto2013}.

A paradigmatic example of a nonequilibrium phase transitions occurs in driven-dissipative nonlinear Kerr oscillators: in the semiclassical limit of large photon numbers and as a function of single-photon drive detuning,  this system  undergoes a first-order transition  manifesting  as a bistability in photon numbers~\cite{Gibbs1976,Drummond1980,Rempe1991, Casteels2016,Casteels2017}. Applying instead a two-photon drive, the resulting Kerr parametric oscillator (KPO) with weak single-photon losses exhibits an additional continuous transition related to the appearance of a parametron which can exist in either of two coherent states of equal amplitude but  $\pi$-phase shifted with respect to each other~\cite{Minganti2016,Elliott2016,Bartolo2016}. At low photon numbers, these coherent states can be recomposed into Schr{\"o}dinger cat states of opposite parities and have been proposed as a new resource for universal quantum computation~\cite{Leghtas2015,Goto2016,Puri2017}. Concurrently, optimization algorithms based on annealing with parametrons have recently been demonstrated using a classical KPO network~\cite{Inagaki2016} with promising quantum extensions~\cite{Nigg2017}.

In this letter, we propose a quantum sensing scheme  based on a first-order symmetry-breaking dissipative phase transition.  This phase transition stems from an explicit breaking of the parity symmetry by the single-photon drive, resulting in an abrupt switching between the coherent states. It is also characterized by a vanishing Liouvillian spectral gap~\cite{Kessler2012}.  
This transition is the quantum manifestation of the classical parametric symmetry breaking studied in Refs.~\cite{Papariello2016, Leuch2016, Eichler2018}. Here, we find that at low and intermediate  photon numbers this switching persists as a sharp crossover. 
Our measurement protocol   extracts  the  {\it unknown} amplitude of an external single-photon drive (signal)  from the detuning frequency at which the KPO  switches  from one coherent state to the other. Remarkably, the switching frequency scales linearly with the amplitude of the single-photon drive, thus realizing a quantum transducer. Furthermore, we discuss the impact of quantum noise on the transducer's sensitivity by simulating a heterodyne detection protocol and by analyzing finite-temperature effects. 
Our results reiterate in a quantum setting the robustness and potential of our detection scheme. Lastly, our scheme is operational in a wide range of parameters, and  readily realizable  in contemporary quantum engineered settings, e.g., in circuit QED, where parametric driving is already utilized for Josephson parametric amplifiers~\cite{Macklin2015}.

\paragraph*{Model.}

\begin{figure}
	\includegraphics[width=1\columnwidth]{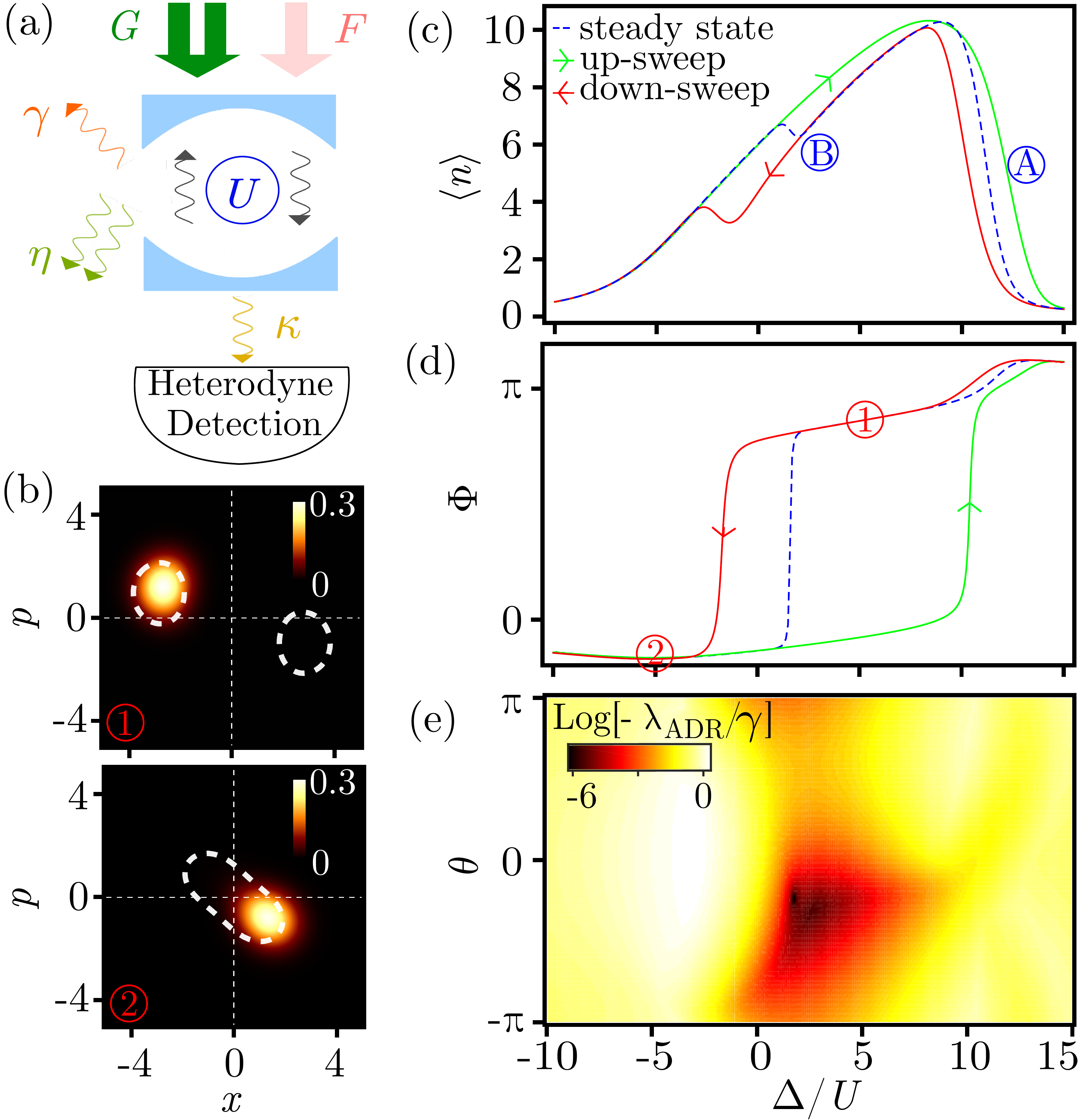}
	\caption{(a) Schematic representation of a Kerr parametric oscillator (KPO) with  nonlinearity $U$ and loss rates $\gamma$ and $\eta$ subject to a single- and two-photon drives $F$ and $G$, respectively. The emitted photons by the cavity with rate $\kappa$ are collected by a heterodyne detector. (b)  Steady-state Husimi $Q$-functions  at points \raisebox{.5pt}{\textcircled{\raisebox{-.9pt} {1}}} and \raisebox{.5pt}{\textcircled{\raisebox{-.9pt} {2}}}, cf.~(c) and (d). Dashed lines mark the contour of the $F=0$ $Q$-function. (c-d) Photon density $\braket{n}$ and phase $\Phi$ of
		the cavity field,  as a function of drive detuning $\Delta/U$ in the steady-state (dashed blue line) and for up-sweeps (green) and down-sweeps (red) of the frequency, obtained from  Eq.~\eqref{eq:lindblad}. 
		At large positive $\Delta/U$,  the KPO crosses over   from high $\braket{n}$ to low $\braket{n}$.  At $\Delta\approx0$ [marked by \raisebox{.5pt}{\textcircled{\raisebox{-.9pt} {B}}}] in (c),  we see a kink in the steady state $\braket{n}$  concomitant with a  $\pi$ switch in the  phase (d).
		The kink and phase switch are also seen for down-sweeps.
		  (e) The  Liouvillian gap as a function of  detuning and phase $\theta$ of the two-photon drive.  The gap vanishes  at the  
		phase-switching transition, marking the onset of a dissipative quantum phase transition. 
		 System parameters are  $F/U = 4$, $|G|/U = 6$, $\gamma/U = 0.5$, $\eta/U = 0.5$.  $\theta=-\frac \pi 2$ in figs. (c) and (d).  The detuning is swept linearly  from $\Delta_1/U = -10$ to  $\Delta_2/U =15$ and vice-versa  in a total sweep time $t_s = 50/U$.
		\label{fig:1} }
\end{figure}

The quantum KPO [Fig.~\ref{fig:1}(a)] is described by the Hamiltonian ($\hbar =1$) 
\begin{equation}
H = -\Delta n + \frac{U}{2} n (n - 1) - (F a + \frac{G^*}{2} a^2 + \text{h.c.})
\label{eq:ham}
\end{equation}
in terms of the bosonic operator $a$ and the number operator $n = a^\dagger a$. The KPO is parametrically pumped with strength 
$G$ while $F$ is the strength of the  single-photon drive; without loss of generality we set $F$ to be real and $G=|G|\exp(i\theta)$. Equation~\eqref{eq:ham} is written in a frame
rotating with respect to the single-photon drive frequency $\omega_d$ and thus the bare cavity frequency is renormalized to the detuning $\Delta=\omega_d-\omega_c$. The parametric modulation is fixed at $2\omega_d$, and $U$ is the Kerr nonlinearity. The dissipative dynamics for the density matrix $\rho$ is determined by the
Lindblad master equation
\begin{equation} 
\label{eq:lindblad}
\dot{\rho} =\mathcal{L} \rho \equiv - i [H,\rho] + \gamma \mathcal{D}[a] \rho + \eta \mathcal{D}[a^2] \rho\,,
\end{equation}
where $\mathcal{L}$ is the Liouvillian superoperator, $\gamma$ and $\eta$ are respectively the single- and two-photon decay rates, and $\mathcal{D}[O] \rho = O \rho O^\dagger - \frac{1}{2} O^\dagger O \rho - \frac{1}{2} \rho O^\dagger O$. 

\paragraph*{Steady-state and dynamics.} When the system is solely subject to a two-photon drive, $F=0$, $G\neq 0$, the system has a $\mathcal{Z}_2$ symmetry associated with the parity operator $P=e^{i\pi a^{\dagger} a}$.  For a wide range of typical experimental parameters, the steady state is given by $\rho_{steady}= c_+ \ket{C_+}\bra{C_+}  + c_- \ket{C_-}\bra{C_-}$ where the cat states $\ket{C_\pm} =c_N (\ket{\alpha} \pm \ket{-\alpha})$, with weighting $c_\pm$, are represented by the coherent states $\ket{\pm \alpha}$ and $c_N$ is a normalization factor~\cite{Minganti2016,Elliott2016}.
 Defining  the Husimi quasi-probability distribution function, $Q(x,p)= \frac{1}{\pi}\braket{x + i p|\rho|x +i p}$, where $\ket{x+ip} \equiv \ket{\alpha}$,  the $\mathcal{Z}_2$ symmetry manifests in the  steady state  as $Q(x,p)=Q(-x,-p)$,  see dashed lines in Fig.~\ref{fig:1}(b).  For $G \gtrsim \gamma,\eta$ and a wide range of detuning around  $\Delta \approx 0$, the Q-function is bimodal  indicating the formation of  cat states. 
For a large enough two-photon drive $G$, the  system is known to  exhibit  both a first-order dissipative phase transition reflecting classical bistability and a continuous  dissipative phase transition  related to the appearance of bimodality in the Husimi Q-function~\cite{Bartolo2016,Minganti2018}.

In the following, we investigate the interplay between the one- and two-photon drives as their detunings are jointly varied.  Since the single-photon drive breaks the $\mathcal{Z}_2$ symmetry,  the  coherent states $\vert\pm \alpha\rangle$ contribute unequally to the   density matrix~\cite{Bartolo2016}. In Fig.~\ref{fig:1}(c), we plot the photon number $\braket{n}$ as a function of  detuning $\Delta/U$. The steady-state photon number  is 
low at large detunings $|\Delta/U| \gg 1$ and increases to  a maximum at $\Delta/U \approx 10$, followed by a  pronounced drop [marked by  \raisebox{.5pt}{\textcircled{\raisebox{-.9pt} {A}}} in Fig.~\ref{fig:1}(c)].    Interestingly, we observe a kink 
occurring at $\Delta/U \approx 0$ [marked by  \raisebox{.5pt}{\textcircled{\raisebox{-.9pt} {B}}}].  This kink  is a precursor to  the  continuous dissipative phase transitions discussed earlier, which is now discontinuous due to the symmetry breaking $F$.

This feature is strongly reflected in the phase of the cavity field $\Phi = \arctan[p/x]$, where $x= \braket{a + a^{\dagger}}$ and $p = \braket{-i (a- a^\dagger)}$. 
 In fact in Fig.~\ref{fig:1}(d), we see that  the  phase  abruptly switches  by $\pi$ in the vicinity of $\Delta/U \approx 0$.  This phase switch  stems directly from the  transition between  the two modes of the parametron in the $Q$-function. Note that these modes are now shifted by the single-photon drive $F$, but nonetheless remain in  opposing  quadrants of the $Q$-function, see Fig.~\ref{fig:1}(b). The origin of this effect can be traced back to the bifurcation physics in the classical limit of the model~\cite{Leuch2016, Papariello2016, Eichler2018}. 

To  substantiate the link between  the phase jump  and dissipative phase transitions, it is
instructive to look at the Liouvillian gap
 $\lambda_{\rm \scriptscriptstyle ADR}$  in the Liouvillian spectrum [Fig.~\ref{fig:1}(e)]. All eigenvalues $\lambda$ of the Liouvillian superoperator $\mathcal{L}$ defined in Eq.~\eqref{eq:lindblad} have negative real parts $\text{Re}(\lambda)\le0$ and we sort them in absolute ascending order $|\text{Re}(\lambda_0)| \leq |\text{Re}(\lambda_1)| \leq ...$~. The lowest eigenvalue $\lambda_0=0$ corresponds to $\rho_{steady}$, and  the Liouvillian gap that determines the slowest decay rate to the steady-state is given by $\lambda_{\rm \scriptscriptstyle ADR} = \text{Re}(\lambda_1)$. The closing of the Liouvillian gap indicates a dissipative  phase transition~\cite{Kessler2012}. 
Our results for 
$\lambda_{\rm \scriptscriptstyle ADR}$ are shown in Fig.~\ref{fig:1}(e), as a function of the  relative driving phase $\theta$ of $G$ and detuning $\Delta/U$.   In the regime where the phase $\Phi$ switches  abruptly,  we find   a vanishingly small Liouvillian gap  $-10^{-6}\gamma$ consistent with the expected first-order transition~\cite{Leuch2016, Papariello2016, Eichler2018}.  Note that for $0 \lesssim \theta \lesssim \pi$ the Liouvillian gap does not close indicating that the phase switching occurs only for $-\pi \lesssim \theta \lesssim 0$.

We now study if the phase switching persists beyond steady state. This is particularly
relevant for  experiments, because the detuning is typically non-adiabatically varied in  time. Simulating the full Lindblad time-evolution 
\eqref{eq:lindblad} under a linear dynamical scan of $\Delta$, we show that both $\braket{n}$ and $\Phi$ manifest a hysteresis cycle, see Figs.~\ref{fig:1}(c) and (d). Such hysteretic behavior survives if the sweep duration is lower than $1/\lambda_{\rm \scriptscriptstyle ADR}$. The steady state  is approached with increasing sweep duration~\cite{supmat}. On the up-sweep only the standard photon number drop at $\Delta\propto G/U$ occurs. Interestingly for down-sweeps, both a marked increase in $\braket{n}$ at $\Delta/U \approx11$ and a kink in $\braket{n}$ concomitant with the phase switching are seen. This is the quantum  analogue of the double hysteresis recently discovered in the classical version of our model~\cite{Leuch2016,  Papariello2016, Eichler2018}. 

\begin{figure*}[t]
	\includegraphics[width=1\linewidth]{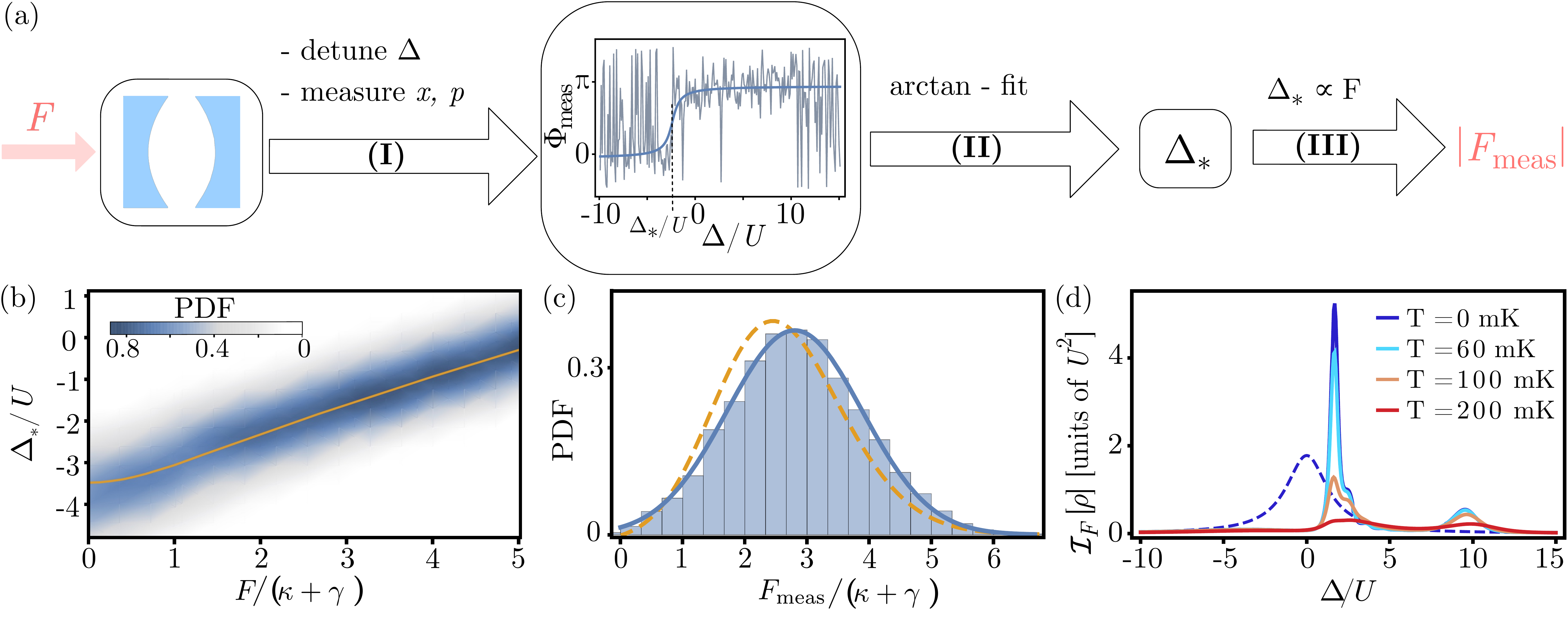}
	\caption{
		(a) Measurement protocol: The phase $\Phi$ is measured via heterodyne detection (grey) and $\Delta_{*}$ is extracted from an $arctan$ fit (blue), which is then used to determine $F_{\rm meas}$ (details see main text).
		(b)  $\Delta_*/U$ as a function of $F$ 
	    obtained from the master equation \eqref{eq:lindblad} (orange line) and its associated  probability density function (PDF, blue) obtained from repeated  heterodyne simulations~\eqref{eq:sme}. 
	    (c)  PDF of $F_{\rm meas}$ for $F/U = 4$, i.e., $F/(\kappa+\gamma) = 2.67$;  histogram from the simulated heterodyne detection (blue) and  a fit to a 	Gaussian distribution function(blue line), with mean $\bar{F}_{\rm meas}/(\kappa+\gamma) = 2.79\approx F/(\kappa+\gamma) $ and standard deviation $\Delta F_{\rm meas}/(\kappa+\gamma) = 1.1$. The dashed line (orange) is  the prediction  based on the stochastic switching in the Husimi-Q function, cf.~Eq.~\eqref{eq:ptrans}. $\kappa/U =1$ and the other parameters  are  the same as in Fig.~\ref{fig:1}; 
		(d) Quantum Fisher information of estimating $F$ in the steady-state~\eqref{eq:QFI} as a function of  $\Delta/U$ (solid lines)~\cite{supmat}. Note the prominent peak at $\Delta=\Delta_*\approx0$ and a smaller one at $\Delta/U \approx 10$, corresponding to the crossovers  \raisebox{.5pt}{\textcircled{\raisebox{-.9pt} {A}}}  and  \raisebox{.5pt}{\textcircled{\raisebox{-.9pt} {B}}}  in Figs.~\ref{fig:1}(c),(d). 
		Both features vanish at sufficiently large temperatures $T$. For comparison, the results for $G=U=0$  for a harmonic oscillator (dashed)
	 at $T=0$ are plotted.
		The main peak is modulated by the resonances in the system \cite{Bartolo2016,Biondi2017}.
		Parameters chosen as $F/U = 4.5$, $|G|/U = 3$, $\gamma/U = 3$, $\eta=0$. 
		\label{fig:2}}
\end{figure*}

The frequency at which the phase switches by $\pi$ for down-sweeps is henceforth labeled by $\Delta_*$.
  We find that, remarkably,  $\Delta_* \propto  F$ over a wide range of single-photon drive amplitudes and relative phases, see Fig.~\ref{fig:2}(b). Departures from this linearity occur when $F$ becomes comparable to 
the loss rates $\gamma$ and $\eta$.   Consubstantial behavior is  seen in the classical limit~\cite{Papariello2016},  but quantum fluctuations entrench the linearity. The linear relation holds for a large range of sweep times, with minor dependences of $\partial\Delta_*/\partial F$ on the sweep time $t_s$~\cite{supmat}. The relation $F\propto\Delta_*$ originating from a phase-switching dissipative phase transition is the key result of our work.  This result can now be exploited to develop a
quantum transducer for measuring forces.

\paragraph*{Quantum transduction protocol.}

To describe a realistic measurement of 
$\Phi$, we simulate continuous observations of $x$ and $p$ as realized in heterodyne detection schemes~\cite{Wiseman2009}. The time evolution of $\rho$ in the presence of the detector can be described by the stochastic master equation 
\begin{align}
\label{eq:sme}
d \rho =& -i [H,\rho] dt + (\gamma+ \kappa) \mathcal{D}[a] \rho dt + \eta \mathcal{D}[a^2] \rho dt  \nonumber \\
&+ \sqrt{\frac{\kappa}{2}} \left( d W_x \mathcal{H} [a] + dW_p \mathcal{H}[-i a] \right) \rho
\end{align}
where $\mathcal{H}[a] \rho = a \rho + \rho a^\dagger - \text{tr}[ a \rho + \rho a^\dagger ]$, $W_{x,p}$ are Wiener processes 
with $\braket{W_i(t)}=0$, $\braket{W_i(t)^2} = t$.
The measurement process effectively increases the single-photon loss rate $\gamma\rightarrow\gamma + 
\kappa$, where $\kappa$ is the emission rate to the heterodyne detector. 
The measured values are given by $x_{\rm meas} = x + \sqrt{2/\kappa}\, dW_x/dt$ and $p_{\rm meas} = p  + 
\sqrt{2/\kappa}\, dW_p/dt$, leading to $\Phi_{\rm meas} = \arctan[p_{\rm meas}/x_{\rm meas}]$. A sample noisy phase measurement is shown in Fig.~\ref{fig:2}(a). 
Our sensing protocol works as follows [Fig.~\ref{fig:2}(a)]: 
(I) the detuning frequency $\Delta(t)$ is varied in a down-sweep and the phase $\Phi_{\rm meas}$ is recorded; (II) to extract $\Delta_*$ from the noisy phase profile, we fit $\Phi_{\rm meas}(\Delta)= \arctan(A(\Delta - \Delta_*) )+C$ with fitting parameters 
$\Delta_*$, $A$, and $C$; (III) the single-photon drive $F_{\rm meas}$ is then obtained using the quasi-linear relation to $\Delta_*$, cf.~orange line in Fig.~\ref{fig:2}(b). Repeating the 
protocol multiple times yields a probability distribution for $\Delta_*$. It matches the result from the averaged master equation \eqref{eq:lindblad}, demonstrating the robustness of our scheme against quantum noise from continuous measurements. Making use of the linear relation, the probability distribution for $\Delta_*$ can be 
translated into a distribution for $F_{\rm meas}$, shown as histograms in Fig.~\ref{fig:2}(c). The distribution can be 
approximated by a Gaussian with standard deviation $\Delta_{F_{\rm meas}}=1.1 (\kappa + \gamma)$ that marks the 
intrinsic quantum noise uncertainty that limits our measurement resolution. The simulations of the heterodyne detection were carried out with QuTiP~\cite{Johansson2013}.

We now show that the PDF obtained from the heterodyne detection can also be determined from the master equation~\eqref{eq:lindblad} with $\gamma \rightarrow \gamma + \kappa$.
Firstly, the Husimi Q-function can be interpreted as a probability density for continuous measurements \cite{Leonhardt:95, Shapiro:84,Stenholm:92,Braunstein:91}. 
As the Q-function changes quadrant across the phase switch at $\Delta=\Delta_*$, [Fig.~\ref{fig:1}(b)],
we introduce the following probabilities
\begin{align}
\label{eq:prob}
\mathcal{P}_{\Phi_-} = \int_{-\infty}^{+\infty} \!\! dp\int_{-\infty}^{0}\!\!dx~Q(x,p) 
\end{align}
and $\mathcal{P}_{\Phi_+}  = 1 - \mathcal{P}_{\Phi_-}$, where $\mathcal{P}_{\Phi_{-(+)}}$ is the probability of measuring the phase in the left (right) half plane. Note that when $\Delta$ is varied in time, the  Husimi Q-function and the corresponding $\mathcal{P}_{\Phi_\pm}$ are time dependent. Let $\mathcal{P}_{m}^i$ denote the probability to measure the phase $m=\Phi_{\pm}$ at time step $i$ and $\mathcal{P}_{m\rightarrow n}^{i\rightarrow i +1}$ the probability to transition from phase  $m=\Phi_\pm$ to phase  $n=\Phi_\mp$ between time steps $i$ and $i+1$.
Making the physically reasonable assumption that the system transitions preferably to the steady state, we obtain the following simple expression for the transition probability to switch from $\Phi_-$ to $\Phi_+$ between the time steps $i$ and $i+1$~\cite{supmat}
\begin{equation}
\label{eq:ptrans}
\mathcal{P}_{\rm tr}^{i\rightarrow i+1} = \mathcal{P}_{\Phi_-}^i - \mathcal{P}_{\Phi_-}^{i+1}.
\end{equation} 
Consequently, for a linear sweep of the detuning, $\mathcal{P}_{\rm tr}(t)
\propto \mathcal{P}_{\rm tr}(\Delta(t)) \equiv \mathcal{P}(\Delta_*=\Delta(t))$. Making use of the linear relation $F \propto \Delta_*$ [Fig.~\ref{fig:2}(b)]  we obtain the PDF of the measured $F$,  $\mathcal{P}(F_{\rm meas}) \propto \mathcal{P}(\Delta_*)$. This simple result qualitatively agrees  with the full PDF obtained from the heterodyne simulation, see 
Fig.~\ref{fig:2}(c).
When $F$ is decreased to very low values, the contributions of both  parametron modes to $\rho$ become comparable and consequently strongly reduces the sensitivity of our  protocol. Moreover in this limit the approach based on Eq.~\ref{eq:ptrans} breaks down.
We note that $F$ is the quantum optical equivalent of a classical mechanically-oscillating force acting on a harmonic oscillator in the rotating-wave approximation~\cite{Ivanov2016}. The measurement protocol discussed here could thus be extended to mechanical forces as well.

\paragraph{Classical noise.} To substantiate the robustness of our proposal, we now  investigate the 
influence of finite temperature  on the  phase-switching in the KPO.  Temperature can induce random switching between the parametron modes, thus potentially  degrading the  fidelity of the sensor.
To quantify this, we  include   an additional dissipative process   in the 
 master equation such that, $\dot{\rho} = - i [H,\rho] + \gamma (1 + n_{\rm th}) 
\mathcal{D}[a] \rho + \gamma n_{\rm th} \mathcal{D}[a^\dagger]\rho$ with $n_{\rm th} = n_{\rm th}(\beta \omega_c)$ 
the thermal number of photons at the  real  frequency of the KPO $\omega_c$, $\beta = k_BT$ and $T$ the temperature of the environment. For simplicity, we 
have neglected two-photon losses ($\eta=0$), since the dominant noise channel is typically 
single-photon loss~\cite{Leghtas2015}.   
A useful  measure to  quantify the sensitivity of our protocol for various temperatures,
is  the quantum Fisher information (QFI).   It is  used to analyze  phase transitions~\cite{Wang2014,Macieszczak2016,Lorenzo2017,Marzolino2017,Irenee2018}  and provides a measure of the variance of  parameter estimations in quantum sensing and metrology~
\cite{Helstrom1976,Braunstein1994}.   Since our sensing  scheme relies on a phase transition, the QFI 
of the steady-state  $\rho$  is particularly appropriate for investigating the role of temperature on the quantum transducer. The QFI quantifies the change of the steady-state density matrix $\rho = \sum_i \lambda_i \ket{\psi_i}\bra{\psi_i}$
w.r.t. variations in the parameter to be estimated, and in our case takes the form
defined as
\begin{equation}
\label{eq:QFI}
\mathcal{I}_F[\rho] = 2 \sum_{ij, \lambda_i + \lambda_j \neq 0} \frac{ |\!\braket{\psi_i| \partial \rho/\partial F |\psi_j}\!|^2}
{\lambda_i + \lambda_j},
\end{equation}
for the estimation of $F$. 

In Fig.~\ref{fig:2}(c), we present the QFI as a function of detuning $\Delta/U$ for 
increasing bath temperatures. We use state-of-the-art KPOs parameters realized in circuit-QED, $\omega_c=7.5\times2\pi$ GHz and $U=25$ kHz~\cite{Leghtas2015}.  We see that
the QFI at $T=0$ (blue) exhibits two sharp peaks 
in correspondence with the crossovers discussed in Fig.~\ref{fig:1}. 
Note that the QFI is largest
 around $\Delta=\Delta_*\sim 0$ where the phase switches, 
while  the  usual bistability transition where the photon number jumps at larger detunings $\Delta/U$, exhibits a lower QFI. 
The QFI of  our  sensing scheme is  therefore, substantially higher  than that of  the standard linear force sensing with the linear oscillator (dashed blue). The QFI  progressively  decreases with temperature, indicating an increasing lower bound for the force estimation variance $\Delta_F$. This bound, however, remains remarkably low for typical operating temperature of circuit-QED devices, $T\approx 20$ mK. This illustrates the potency of our sensing protocol based on a dissipative phase transition  for sensitive  measurements.

\paragraph*{Outlook.} 
We have proposed a quantum sensing scheme that relies on the heightened sensitivity of driven-dissipative phase transitions.  Our transduction scheme is widely-realizable in contemporary quantum engineered devices, including optical~\cite{Hartmann2016,Noh2016}, mechanical~\cite{Rossi2018}, and electronic~\cite{Schmidt2013,Leghtas2015} platforms. A key ingredient for our proposal relies on the control of single- and two-photon drives, which are readily accessible in such systems using standard nonlinear wave-mixing techniques~\cite{Shen1984}. Our work opens interesting perspectives in studying the interplay of sensing and entanglement in networks of KPOs vis-a-vis synchronization and other collective many-body effects~\cite{Lee2014, Savona2017, Biondi2015, Baboux2015}.

\begin{acknowledgments}
We thank L. Papariello and A. Eichler for fruitful discussions. We acknowledge financial support from the Swiss National Science Foundation and the Sinergia grant CRSII5\_177198.
\end{acknowledgments}

\newpage
\cleardoublepage
\setcounter{figure}{0}
\renewcommand{\figurename}{Supplemental Material, Figure}

\onecolumngrid
\begin{center}
	\textbf{\normalsize Supplemental Material for}\\
	\vspace{3mm}
	\textbf{\large A quantum transducer using a parametric driven-dissipative phase transition}
	\vspace{4mm}
	
	{ Toni L. Heugel, Matteo Biondi, Oded Zilberberg, and R. Chitra}\\
	\vspace{1mm}
	\textit{Institute for Theoretical Physics, ETH Zurich, 8093 Z\"urich, Switzerland
	}

	\vspace{5mm}
\end{center}
\setcounter{equation}{0}
\setcounter{section}{0}
\setcounter{figure}{0}
\setcounter{table}{0}
\setcounter{page}{1}
\makeatletter
\renewcommand{\bibnumfmt}[1]{[#1]}
\renewcommand{\citenumfont}[1]{#1}
\renewcommand{\theequation}{S\arabic{equation}}
\renewcommand{\thefigure}{S\arabic{figure}}
\twocolumngrid

\section{Numerics}\label{sec:1}
The results in the main text for the steady state, up- and down-sweeps of  the detuning, as well as  heterodyne detection were obtained  by numerically solving the corresponding master equations. The master equation~[Eq.~(2) in the main text] is solved in the Fock basis of the resonator. For this, we represent the density matrix in a   truncated  Fock basis  of  $N$  states and neglect all contributions from other Fock states.  We explicitly check for  the convergence of our results as a function of $N$.
In the Fock basis, since the density matrix can be rewritten as a column vector and the Liouvillian as a matrix, the master equation reduces to a  set of coupled differential equations which can be solved using standard numerical packages.
The steady state is found as the eigenstate of the Liouvillian matrix corresponding to the eigenvalue $0$, while the dynamical $\Delta$ sweeps are simulated by numerically integrating the ordinary differential equation. The heterodyne detection  scheme used to discuss a measurement of the phase was was simulated using QuTiP~\cite{Johansson2013}.

\section{Dependence  of $\Delta_*(F)$ on the sweep time}

\begin{figure}[!h]
	\includegraphics[width=0.9\linewidth]{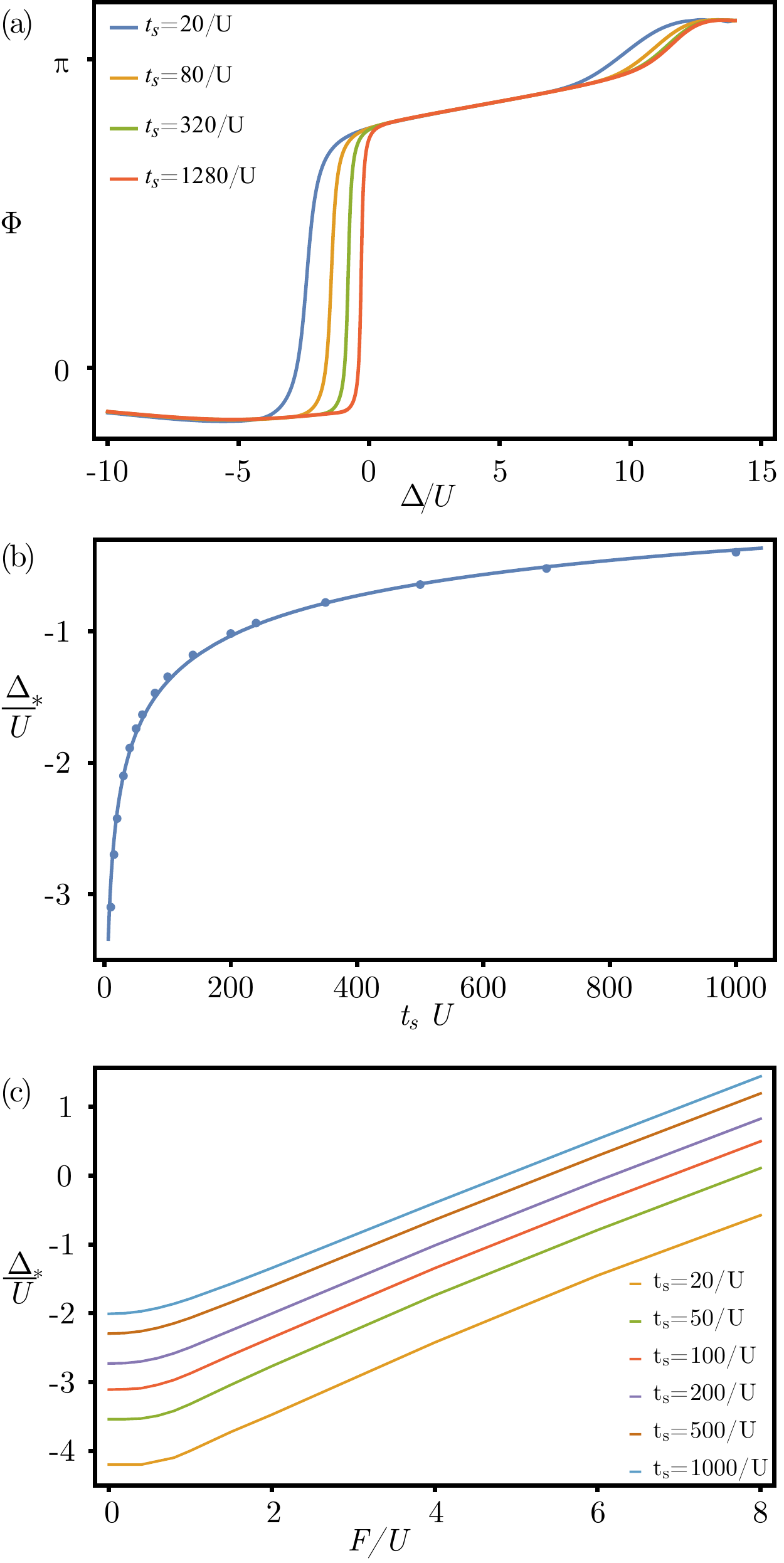}\\
	\caption{\label{fig:Tsweep}(a) Phase $\Phi$ of the cavity field as a function of $\Delta$ for different sweep times $t_s$ (increasing from left to right). (b) $\Delta_*$ as a function of the sweep time $t_s$ showing the convergence. The points are from simulations and the line is a fit of $\frac{b}{x^a} +c$, where $c$ gives the steady state value of $\Delta_*$, $a=0.18$ and $b=-3.91$. (c) $\Delta_*$ as a function of $F$ for different sweep times $t_s$ (increasing from below). Parameter values: $F/U = 4$, $G/U = 6$, $\gamma/U=0.5$ and $\eta/U = 0.5$.}
\end{figure}

The rate of the frequency sweeps directly affects the detuning $\Delta_*$ where the phase $\Phi$ switches. As the duration of the sweep $t_s$ is increased, the photon number and the phase of the up- and down-sweep approach the results of the steady state. Here, we analyze this effect quantitatively. We only consider the down-sweep since it determines $\Delta_*$. In Fig.~\ref{fig:Tsweep}(a), we plot  the phase as function of $\Delta$ for different sweep times $t_s$. The larger  the
sweep time $t_s$, the larger the $\Delta_*$ and the steeper  the  switch between the two coherent states. In Fig.~\ref{fig:Tsweep}(b),  we present $\Delta_*$ as a function of $t_s$. We find that the function $\frac{b}{x^a} +c$ fits the curve, where $c$ gives the steady state value of $\Delta_*$. The fit yields $a=0.18$ and $b=-3.91$. In Fig.~\ref{fig:Tsweep} (c), $\Delta_*$ as a function of $F$ is depicted for different values of $t_s$. We find that the convergence behavior of $\Delta_*$ is almost independent of the applied coherent drive $F$,  indicating  that the shape of $\Delta_*(F)$ does not change with $t_s$. Larger $t_s$ only shifts $\Delta_*(F)$ to larger values and  has minimal impact on the 
slope.

\section{Dependence on  the dissipative coefficients $\kappa/\gamma$}

\begin{figure}
	\centering
	\includegraphics[width=0.9\linewidth]{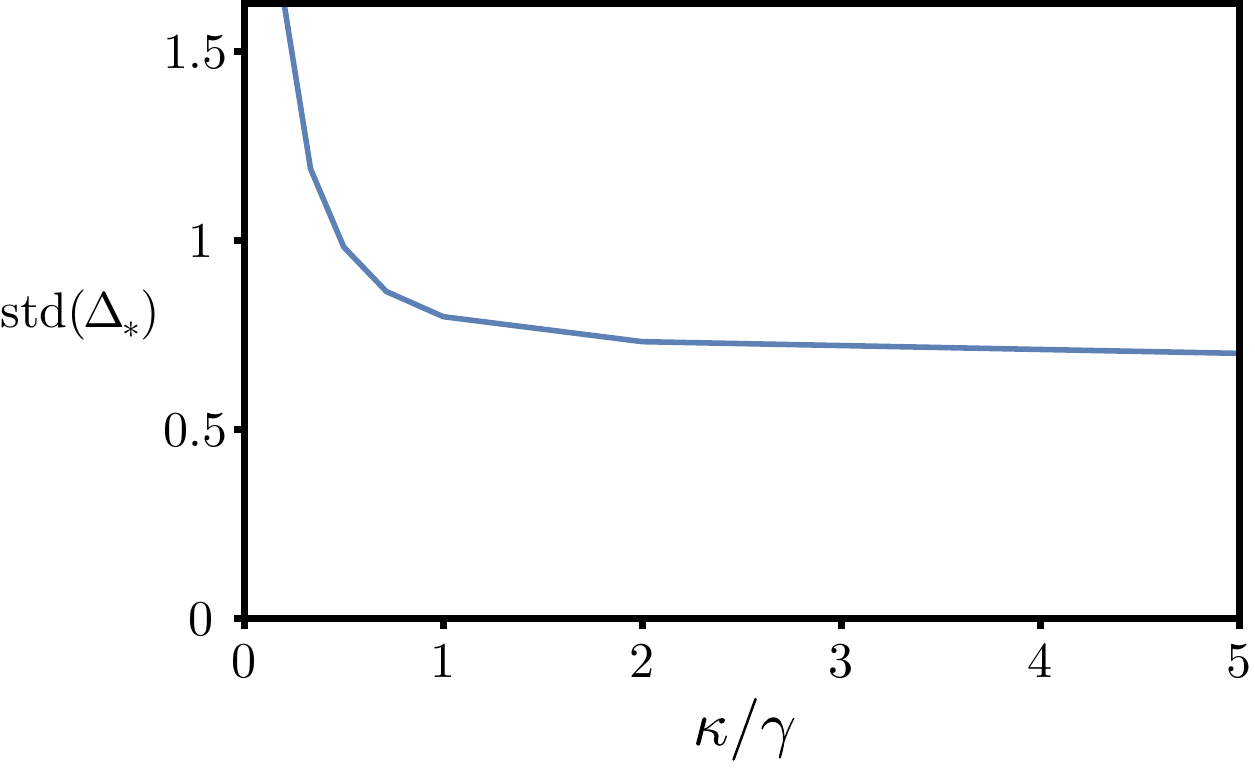}
	\caption{\label{fig:kappagamma}Standard deviation (std) of the measured $\Delta_*/U$ as a function of $\kappa/\gamma$, where $\kappa + \gamma = 1.5U$ is fixed. Other parameter values: $F/U = 4$, $G/U = 6$, $\eta/U = 0.5$ and $t_s=50/U$.}
\end{figure}

Next, we study the dependence of the standard deviation of $\Delta_*/U$ on $\kappa/\gamma$, when $(\kappa + \gamma)/U = 1.5$. In Fig.~\ref{fig:kappagamma}, we show that the standard deviation is monotonously decreased with $\kappa/\gamma$. For $\kappa \rightarrow 1.5$ and $\gamma\rightarrow 0$, the standard deviation converges to $0.67$. As this maximizes the detection rate the error in the measured phase, $\Phi_{\rm{meas}}$ is minimized while the dissipation is fixed ($(\kappa + \gamma)/U = 1.5$). Therefore, the standard deviation in $\Delta_*/U$ is minimal.  As $\kappa$ goes to $0$ the standard deviation goes to infinity, since the fluctuations in $x_{\rm{meas}}$ and $y_{\rm{meas}}$ are proportional to $1/\sqrt{\kappa}$ (see main text). 

\section{Derivation of the transition probability}
In this section, we present a derivation of Eq.~(5) in the main text.
We have introduced the probabilities
\begin{align}
	\mathcal{P}_{\Phi_-} = \int_{-\infty}^{+\infty} \!\! dp\int_{-\infty}^{0}\!\!dx~Q(x,p) \,,
\end{align}
and $\mathcal{P}_{\Phi_+}  = 1 - \mathcal{P}_{\Phi_-}$, where $\mathcal{P}_{\Phi_{-(+)}}$ gives the probability of measuring the phase in the left (right) half plane. The probabilities $\mathcal{P}_{\Phi_\pm}$ are time-dependent functions. Let $\mathcal{P}_{m}^i$ denote the probability to measure the phase $m=\Phi_{\pm}$ at time step $i$ and $\mathcal{P}_{m\rightarrow n}^{i\rightarrow i +1}$ the probability to transition from phase  $m=\Phi_\pm$ to phase  $n=\Phi_\mp$ between time steps $i$ and $i+1$.
The probability to find the system in the state $m$ at time step $i+1$ is then given by
\begin{equation}
	\mathcal{P}_{m}^{i+1} = \mathcal{P}_{m}^i \mathcal{P}_{m\rightarrow m}^{i\rightarrow i +1} + \mathcal{P}_{n}^i \mathcal{P}_{n\rightarrow m}^{i\rightarrow i +1}\,.
\end{equation}
Combing these equations and using $1 = \mathcal{P}_{n\rightarrow m}^{i\rightarrow i +1}+\mathcal{P}_{n\rightarrow n}^{i\rightarrow i +1}$, we obtain $\mathcal{P}_{n}^{i+1} - \mathcal{P}_{m}^{i+1} = \mathcal{P}_{n}^{i} - \mathcal{P}_{m}^{i} + 2 \left( \mathcal{P}_{m}^i \mathcal{P}_{m\rightarrow n}^{i\rightarrow i +1} -\mathcal{P}_{n}^i \mathcal{P}_{n\rightarrow m}^{i\rightarrow i +1} \right)$. Next, we assume that $\mathcal{P}_{\Phi_+}^i \mathcal{P}_{\Phi_+ \rightarrow \Phi_-}^{i\rightarrow i +1}$ can be neglected since the steady state at the transition is at $\Phi_+$, while the state of the dynamic evolution transitions from $\Phi_-$ to $\Phi_+$.
The transition probability $
\mathcal{P}_{\rm tr}$ to switch from $\Phi_-$ to $\Phi_+$ between the time steps $i$ and $i+1$ is described by
\begin{equation}
	\mathcal{P}_{\rm tr}^{i\rightarrow i+1} = \mathcal{P}_{\Phi_-}^i \mathcal{P}_{\Phi_-\rightarrow \Phi_+}^{i\rightarrow i +1} = \mathcal{P}_{\Phi_-}^i - \mathcal{P}_{\Phi_-}^{i+1}
\end{equation} 
as $\Delta(t)$ is down-swept from $\Delta_1/U=15$ at $t=t_s$ 
to $\Delta_2=-10/U$ at $t=2t_s$ across the phase switch.

\section{Quantum Fisher Information}
In the main text, the quantum Fisher information (QFI) is used to study the effect of temperature on the sensing scheme. In order to calculate the QFI one needs to diagonalize the density matrix $\rho$, see Eq.~(6) in the main text. For dissipative systems the steady state is often given by a mixed state and therefore the analytical diagonalization can be difficult. However,  pure states $\rho = \ket{\psi}\bra{\psi}$ are already diagonal and the QFI simplifies to 
\begin{equation}
	\mathcal{I}_F[\rho] = 4\left( \braket{\partial_F \psi | \partial_F \psi} - |\braket{\psi | \partial_F\psi} |^2 \right),
\end{equation}
where $\ket{\partial_F\psi} = \frac{\partial}{\partial F} \ket{\psi}$.

Firstly, we look at the linear case ($U=0$, $G=0$ and $\eta=0$) at $T=0$, where the QFI of the steady state $\mathcal{I}_F[\rho]$ can be calculated analytically.  The steady state solution ($\dot{\rho} = 0$) of Eq.~(2) is given by the pure state $\rho = \ket{\alpha}\bra{\alpha}$, with coherent state
\begin{equation}
	\ket{\alpha} = \ket{\frac{i F}{\frac{\gamma}{2} - i \Delta}}.
\end{equation}
From this we  find an analytical expression for the  QFI
\begin{equation}
	\mathcal{I}_F[\rho] = \frac{4}{F^2}  \left( |\alpha|^2 + 2 |\alpha|^4 -|\alpha|^4 -|\alpha|^4 \right) = \frac{4}{\frac{\gamma^2}{4} + \Delta^2},
\end{equation}
where we used $\partial_F \ket{\alpha} = \frac{1}{F} (\alpha a^\dagger - \alpha^* a) \ket{\alpha}$.

In the nonlinear case ($U\neq0$, $G\neq0$ and $\eta\neq0$) the Fisher information can be calculated numerically from the density matrix $\rho$. As discussed in section~\ref{sec:1}, the state $\rho$ is calculated in a truncated Fock basis.  Thus, we can diagonalize $\rho$ and calculate the derivatives numerically. The QFI is then determined using Eq.~(6) in the main text or with~\cite{Salvatori:14}
\begin{equation}
	\mathcal{I}_F[\rho] = \sum_{i} \frac{(\partial_F p_i)^2}{p_i} + 2 \sum_{i\neq j, p_i+p_j\neq 0} \frac{(p_i - p_j)^2}{p_i + p_j} | \braket{\psi_i|\partial_F \psi_j}|^2. 
\end{equation}

\end{document}